\documentclass[conference]{IEEEtran}
\IEEEoverridecommandlockouts

\usepackage{amsmath}
\usepackage{amssymb} 
\usepackage{amsthm}
\usepackage{mathtools}
\usepackage[caption=false, font=footnotesize]{subfig}
\usepackage{graphicx}
\usepackage{xcolor}
\usepackage{algorithm,algorithmic}

\theoremstyle{definition}

\theoremstyle{remark}
\newtheorem{remark}{Remark}

\theoremstyle{definition}

\theoremstyle{definition}

\theoremstyle{definition}

\title{\LARGE \bf Q-learning Based Optimal False Data Injection Attack on Probabilistic Boolean Control Networks} 

\author{Xianlun Peng, Yang Tang,  Fangfei Li and Yang Liu
\thanks{This research is funded by the National Natural Science Foundation of China under Grants 62173142, 62173308, the Programme of Introducing Talents of Discipline to Universities (the 111 Project) under Grant B17017 and Zhejiang Provincial Natural Science Foundation of China under Grants LR20F030001, LD19A010001.}%
\thanks{Xianlun Peng is with School of Mathematics, East China University of Science and Technology, Shanghai 200237, China.}
\thanks{Yang Tang is with the Key Laboratory of Smart Manufacturing in Energy Chemical Process, Ministry of Education, East China University of Science and Technology, Shanghai 200237, China.}
\thanks{Fangfei Li is with School of Mathematics and the Key Laboratory of Smart Manufacturing in Energy Chemical Process, Ministry of Education, East China University of Science and Technology, Shanghai 200237, China.}
\thanks{Yang Liu is with the Key Laboratory of Intelligent Education Technology and Application of Zhejiang Province, College of Mathematics and Computer Science, Zhejiang Normal University.}

}

\begin{document}

\thispagestyle{empty}
\pagestyle{empty}
\maketitle

\begin{abstract}                          
    In this paper, we present a reinforcement learning (RL) method for solving optimal false data injection attack problems in probabilistic Boolean control networks (PBCNs) where the attacker lacks knowledge of the system model. Specifically, we employ a Q-learning (QL) algorithm to address this problem. We then propose an improved QL algorithm that not only enhances learning efficiency but also obtains optimal attack strategies for large-scale PBCNs that the standard QL algorithm cannot handle. Finally, we verify the effectiveness of our proposed approach by considering two attacked PBCNs, including a 10-node network and a 28-node network.
\end{abstract}

\section{Introduction}
Boolean networks (BNs) were initially proposed by Kauffman \cite{kauffman1969metabolic} as a discrete system based on
directed graphs. In BNs, each node can only be in one of two state values at a given time,
namely on (1) or off (0). The state of each node at the next moment is updated by a series of
logical functions \cite{Mengm2019tac64790,RobustnessbooleanstochasticTAC}. When control inputs are considered, BNs are called Boolean control
networks (BCNs). BNs have received considerable attention and have been widely used in
many fields, including cell differentiation, immune response, biological evolution, neural
networks, and gene regulation \cite{BooleaKF2020Auto111108609,heidel2003finding,Guo2015Set}. Probabilistic Boolean control networks (PBCNs) were
proposed by \cite{shmulevich2002probabilistic} as meaningful extensions of classic BCNs. In PBCNs, each node corresponds to one
or more Boolean functions, and one of these functions is selected with a certain probability
at each iteration step to update its next state. PBCNs have a wide range of applications,
including biological systems \cite{ma2008probabilistic}, manufacturing engineering systems \cite{rivera2018probabilistic1}, biomedicine \cite{trairatphisan2013recent}, credit
defaults \cite{gu2013modeling}, and industrial machine systems \cite{rivera2018probabilistic}, etc.

In recent years, the rapid development of information technology has attracted scholarly
attention to cyber-physical systems (CPSs) \cite{2015Deterministic}. CPSs use computing and communication cores to monitor, coordinate, control, and integrate physical processes. They have a wide range of
applications, including smart buildings, smart grids, and intelligent transportation \cite{ahmed2013cyber,TangTCBinpress}. Recently,
\cite{RoliA2011ECAEC43} proposed using BNs to describe a type of CPS called Boolean network robots. Similar to
the transmission of general CPSs, the control signals of the Boolean network robots
proposed in \cite{RoliA2011ECAEC43} are transmitted to the actuator via wireless networks. However, because the
control inputs of CPSs are transmitted through an unprotected wireless network, they are
vulnerable to cyber attacks \cite{optimalattackroundrobinzhangJ}. For example, Stuxnet is a malware that tampers with the transmission data of industrial control systems, causing significant loss of system
performance \cite{Markoff2010NYT1601}. Recently, there has been increasing research interest in the cyber security of CPSs and many results have been published on the impact of malicious attacks \cite{optimalattackroundrobinsuiT}.

Attack models can generally be divided into three types: 1) denial-of-service (DoS) attacks, 2)
replay attacks, and 3) false data injection attacks. DoS attacks jam communication channels,
preventing the target system from providing normal services or accessing resources \cite{ZhangH2015IEEETAC603023}. Replay attacks involve an attacker accessing, recording, and replaying sensor data \cite{ChenBIEEETCB}. False data injection attacks, first introduced in power networks, involve an attacker modifying sensor data to destabilize the system \cite{LiuY200916ACM21}. Since the statistical characteristics of the data usually do not change when replay and false data injection attacks are launched, it is difficult for the system to detect them, resulting in significant losses. There have been many results on replay and
false data injection attacks in power systems, SCADA systems, etc \cite{mo2013detecting,yang2019distributed}.

To our knowledge, although some results have been achieved on the security of CPSs, the
attack problem of PBCNs, as a class of CPSs, has not been reported. Only by understanding
possible attack strategies from the perspective of attackers can we provide effective policies
for defending the system. Therefore, studying the attack problem of a PBCN is very
meaningful. Additionally, the current attack problems for CPSs typically arise when the attacker possesses in-depth knowledge of the system model, while there is limited research available when the attacker lacks knowledge of the system.

Reinforcement Learning (RL) is a paradigm and methodology of machine learning applicable
to systems with unknown models. It is used to describe and solve problems where intelligent
agents learn strategies to maximize benefits or achieve specific goals while interacting with
their environment. A common model of RL is the Markov Decision Process (MDP).
Depending on given conditions, RL can be classified as model-based or model-free, and has
been widely studied in the literature \cite{sutton2018reinforcement,bucsoniu2018reinforcement,preitl2007iterative,roman2021hybrid}. The Q-learning (QL) algorithm, a model-free method
for solving RL problems, was first proposed by Watkins in 1989 \cite{watkins1989learning}. The QL algorithm has been
used to address control problems in model-free Boolean networks \cite{liu2021weak}. However, although some
interesting results have been reported on the control problems of model-free PBCNs, to the
best of our knowledge, the attack problems of PBCNs, especially in the case of model-free,
have not been solved. Additionally, the QL algorithm can only be used to address control
problems in small-scale Boolean networks. Although scholars have used deep Q networks
(DQNs) to study control problems in large-scale Boolean networks, their computational
complexity is higher than that of QL and, unlike QL, the convergence of DQNs cannot be
guaranteed. Therefore, it is of great significance to propose an attack strategy for PBCNs that
is not only applicable to model-free situations but also to large-scale PBCNs.

Motivated by the above analysis, we consider the optimal false data injection attack on a
PBCN where the attacker has no information about the system model. We use an MDP to
model this problem and provide the optimal attack strategy from the perspective of attackers
based on QL. We also modify the QL algorithm to improve learning efficiency and make it
applicable to the optimal attack problem of some large-scale PBCNs. The main contributions
of this paper are as follows:

\begin{enumerate}
    \item We present the dynamics of false data injection attacks on PBCNs within the MDP framework
    and employ the QL algorithm (Algorithm 1) to solve three problems of interest regarding the
    optimal false data injection attack on PBCNs, where the attacker has no knowledge of the
    system model.
    \item We propose Algorithm 2 to improve the QL algorithm by changing the way action-value
    functions are stored, which can dramatically reduce the space requirements of these
    functions. As a result, it can solve the optimal attack problem for some large-scale PBCNs
    that the standard QL algorithm cannot handle. 
    \item Additionally, by recording the expected total
    reward of a policy during the training process to avoid unnecessary learning processes,
    Algorithm 2 can improve learning efficiency.
\end{enumerate}

The rest of the paper is organized as follows. Section 2 gives a brief introduction of MDP. In section 3, we introduce the PBCN and give the model of the false data injection attack for a PBCN in the MDP framework. In section 4, We propose a QL algorithm, an improved QL algorithm, which cannot only deal with the optimal attack problem for PBCNs, but also for some large scale PBCNs effectively. Finally, in section 5, we use two examples to illustrate the effectiveness of the proposed algorithms.  

{\bfseries Notations:\;}$\mathbb{R}$ denotes the sets of real numbers. Let $\mathcal{B} \doteq \{0,1\}$, $\mathcal{B}^{n} \doteq \underbrace{\mathcal{B} \times \ldots \times \mathcal{B}}\limits_{n}$. Denote the basic logical operators ``Negation",``And", ``Or" by $\neg$, $\land$, $\vee$, respectively.

\section{Markov Decision Process}
In this section, we give a brief introduction of MDP that will be used in the sequel.
A discrete-time Markov decision process (MDP) \cite{dimitri1995dynamic} is a quadruple $\mathcal{(S,A,R,P)}$, where $\mathcal{S}$ is the state-space, $\mathcal{A}$ is the action-space, $\mathcal{R}$ denotes the rewards, and $\mathcal{P}:\mathcal{S \times R \times S \times A \to} [0,1]$ is the function that defines the dynamics of the MDP. That is, for each state $S_{t-1} \in \mathcal{S}$ and action $A_{t-1} \in \mathcal{A}$, the following transition probability describes the conditional probability of transition from  $S_{t-1}$ to $S_{t}$ when action $A_{t-1}$ is taken:
\begin{equation}
    \mathcal{P}(s^{\prime},r | s,a) \doteq Pr\{S_t=s^{\prime},R_t=r | S_{t-1}=s,A_{t-1}=a\}.
\end{equation}
Moreover, let $R_{t+1} \in  \mathcal{R}$ denotes the reward resulting from $S_{t+1}, A_{t}$, and $S_{t}$. Informally, the goal of the RL is to maximize the total amount of the reward. Define the sum of future reward at time-step $t$:
\begin{equation}
    G_t \doteq \sum_{k=t+1}^{T}\gamma^{k-t-1}R_k
\end{equation}
where $T$ is the final time step, $0 \le \gamma \le 1$ is the discount rate, and including the possibility that $T = \infty$ or $\gamma = 1$ (but not both). The basic RL problem is to find a policy $\pi$ that maximizes the expectation of $G_t$, i.e., $ \mathbb{E}_{\pi}[G_t] $, at each $t$. Formally, a policy is a mapping from states to probabilities of selecting each possible action, and we use $\pi^*$ to denote the optimal policy.

The value function when the agent starts in the state  $s$ under the policy $\pi$, which is denoted by $v_{\pi}(s)$, is given as follows:

\begin{equation}
    v_{\pi}(s) \doteq \mathbb{E}_{\pi}[G_t | S_t=s] = \mathbb{E}_{\pi}[\sum_{k=t+1}^{T}\gamma^{k-t-1}R_k | S_t=s] 
\end{equation}
for all $s \in \mathcal{S}$. Similarly, the value of taking action $a$ in state $s$ under a policy $\pi$, which is denoted by  $q_{\pi}(s,a)$, is given as follows:

\begin{equation}
    \begin{split}
        q_{\pi}(s,a) & \doteq \mathbb{E}_{\pi}[G_t | S_t=s, A_t=a] \\
    & =\mathbb{E}_{\pi}[\sum_{k=t+1}^{T}\gamma^{k-t-1}R_k | S_t=s, A_t=a]
    \end{split}
\end{equation}
for all  $s \in \mathcal{S}$, $a \in \mathcal{A}$. In addition, we call the function $q_{\pi}$ the action-value function for policy $\pi$.

A fundamental property of the value functions $v_{\pi}(s)$ and $q_{\pi}(s,a)$ is that they satisfy the recursive Bellman equations:
\begin{equation}
    v_{\pi}(s) = \sum_{a}\pi(a|s)\sum_{s^{\prime},r}\mathcal{P}(s^{\prime},r|s,a) \big [r+\gamma v_{\pi}(s^{\prime}) \big ]
\end{equation}
for all $s \in \mathcal{S}$, and:
\begin{equation}
    q_{\pi}(s,a) = \sum_{s^{\prime},r}\mathcal{P}(s^{\prime},r|s,a) \big [r+\gamma \sum_{a^{\prime}}\pi(a^{\prime}|s^{\prime})q_{\pi}(s^{\prime},a^{\prime}) \big ]
\end{equation}

for all $s \in \mathcal{S}$ and $a \in \mathcal{A}$, where $\pi(a|s)$ represents the conditional
probability of taking the action $A_t=a$ if state $S_t=s$.

For finite MDPs, an optimal policy can be defined by a partial ordering over policies. That is, $\pi \geq \pi^{\prime}$ if and only if $v_{\pi}(s) \geq v_{\pi^{\prime}}(s)$ for all $s \in \mathcal{S}$. Then, the optimal policy denoted by $\pi_*$
can be defined as:
$$\pi_{*} \doteq \mathop{\rm argmax}\limits_{\pi \in \prod} v_{\pi}(s), {\rm for \ all} \ s \in \mathcal{S}$$
where $\prod$ is the set of all admissible policies. It should be pointed out that there may be more than one optimal policy, and they share the same state-value function, called the optimal state-value function, which is defined as:
$$v_{*}(s) \doteq \mathop{{\rm max}}\limits_{\pi}v_{\pi}(s)$$
for all $s \in \mathcal{S}$. In addition, the optimal policies also share the same optimal action-value function, which is defined as:
$$q_{*}(s,a) \doteq \mathop{{\rm max}}\limits_{\pi}q_{\pi}(s,a)$$
for all $s \in \mathcal{S}$, $a \in \mathcal{A}$. For $v_{*}$ and $q_{*}$, there is an equation similar to Bellman equation, called Bellman optimality equation:
\begin{equation}
    \begin{split}
        v_{*}(s) & = \mathop{{\rm max}}\limits_{a \in \mathcal{A}(s)}q_{*}(s,a) \\
        & = \mathop{{\rm max}}\limits_{a}\mathbb{E}_{\pi_{*}} \big [ R_{t+1}+\gamma v_{*}(S_{t+1})|S_{t}=s,A_{t}=a \big ] \\
        & = \mathop{{\rm max}}\limits_{a}\sum_{s^{\prime},r}\mathcal{P}(s^{\prime},r|s,a)[r+\gamma  v_{*}(s^{\prime})]
    \end{split}
\end{equation}
for all $s \in \mathcal{S}$, and:
\begin{equation}
    \begin{split}
        q_{*}(s,a) & = \mathbb{E} \big [ R_{t+1}+ \gamma \mathop{{\rm max}}\limits_{a^{\prime}}q_{*}(S_{t+1},a^{\prime}) | S_t=s, A_t=a \big ] \\
    & = \sum_{s^{\prime},r}\mathcal{P}(s^{\prime},r|s,a) \big [ r+\gamma \mathop{{\rm max}}\limits_{a^{\prime}}q_{*}(S^{\prime},a^{\prime})) \big ]
    \end{split}
\end{equation}
for all $s \in \mathcal{S}$, $a \in \mathcal{A}$. When the 
optimal state-value function or the optimal action-value function is given, one can obtain an optimal deterministic policy as:
$$\pi_{*}(s)=\mathop{{\rm argmax}}\limits_{a}\sum_{r,s^{\prime}}\mathcal{P}(s^{\prime},r|s,a)[r+\gamma v_{*}(s^{\prime})]$$
or:
$$\pi_{*}(s)=\mathop{{\rm argmax}}\limits_{a}q_{*}(s,a)$$
for all $s \in \mathcal{S}$.

\section{System and Attack Model}
\label{sec:guidelines}
In this section, we introduce the PBCN model. We assume that there is a malicious attacker that tampers with the control inputs of the system in order to change the steady state of the system. 

\subsection{Probabilistic Boolean Control Networks}
A PBCN with $n$ nodes, $m$ control inputs can be described as:
\begin{equation}
    \begin{cases}
 x_{1}(t+1)= f_{1}(u_{1}(t), \ldots, u_{m}(t), x_{1}(t), \ldots, x_{n}(t)) \\ 
 x_{2}(t+1)= f_{2}(u_{1}(t), \ldots, u_{m}(t), x_{1}(t), \ldots, x_{n}(t)) \\
 \vdots \\
 x_{n}(t+1)= f_{n}(u_{1}(t), \ldots, u_{m}(t), x_{1}(t), \ldots, x_{n}(t))
\end{cases} 
\end{equation}
where $x_{i}(t) \in \mathcal{B}$ is the state of node $i$, $u_{i}(t) \in \mathcal{B}$ is the $i$th control input. In addition, $f_{i} \in \{ f_{i}^{1}, f_{i}^{2}, \ldots, f_{i}^{l_{i}} \}: \mathcal{B}^{n+m} \rightarrow \mathcal{B}, i=1, 2, \ldots, n$, are logical functions randomly chosen with probabilities $\{ p_{i}^{1}, p_{i}^{2}, \ldots, p_{i}^{l_{i}} \} $, where ${l_{i}}$ is the number of logical functions that can be chosen by node $i$, $\sum_{j=1}^{l_{i}}p_{i}^{j}=1 $ and $p_{i}^{j} \geq 0 $. The PBCN (9) becomes a BCN when $l_{i}=1$, for $i=1, 2, \ldots, n$. We assume that the selection for each logical function $f_{i}$ is independent, i.e.
$$Pr\{f_{i}=f_{i}^{j}, f_{l}=f_{l}^{k}\}=Pr\{f_{i}=f_{i}^{j}\} \cdot Pr\{f_{l}=f_{l}^{k}\}=p_{i}^{j} \cdot p_{l}^{k}. $$
Given an initial state $X(0) \doteq (x_{1}(0),x_{2}(0),\ldots,x_{n}(0)) \in \mathcal{B}^{n}$ and a sequence of controls $U_t \doteq \{U(0),U(1),\ldots,U(t-1)\}$, where $U(i)\doteq (u_{1}(i),u_{2}(i),\ldots,u_{m}(i)) \in \mathcal{B}^{m}$, the solution of the PBCN (9) is denoted by $X(t;X(0);U_t)$. If there exists a control $U_1$ such that $Pr\{X(1;X(0);U_1)=X(0)\}=1$, then we call state $X(0)$ the equilibrium point. Note that there may be more than one equilibrium point.

\subsection{Attacked PBCN Model}
As we discussed in the introduction, when we use a PBCN to model a CPS, the control signal is transmitted to the actuator side through wireless network, which is vulnerable to cyber attack in many cases. Therefore, in this section, we consider an attacker who attacks the wireless network and tampers with the control signals, aiming to change the system state.
\begin{figure}[!htbp]
\centerline{\includegraphics[width=\columnwidth]{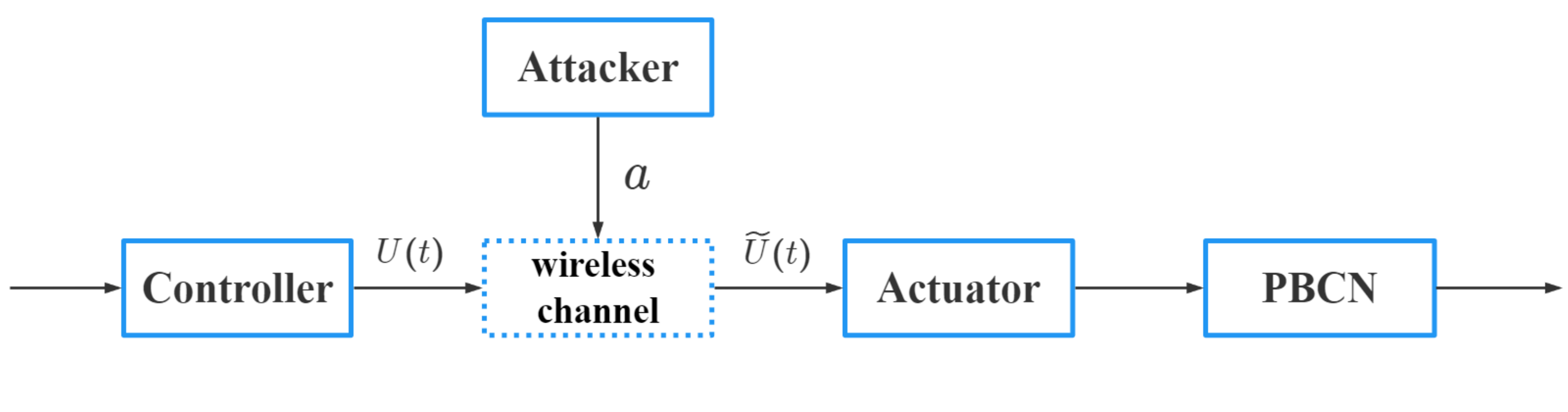}}
\caption{The attack framework.}
\label{fig1}
\end{figure}

Assume that the PBCN is initially in the state $X^{*}$ at time $t=0$, denoted as $X(0)=X^{*}$ and it reaches the state $X(T)$ at time $t=T$ under a sequence of controls $U_{T}$ with probability 1. Considering that the attacker can attack any control node and invert the value of the control node, the attack framework is shown in Figure 1. We define the  possible actions at each time step as $a(t)\doteq(a_1(t),a_2(t),\ldots,a_m(t))\in \mathcal{B}^{m}$, where $a_{i}(t) \in \mathcal{B},\; i=1,2,\ldots,m$ and $a_{i}=0$ for no attack, $a_{i}=1$ for attack on the $i$th control node. Then, the number of actions at each state is $2^m$, and the control $U(t)$ is changed to $\widetilde{U}(t)=(\widetilde{u}_{1}(t),\widetilde{u}_{2}(t),\ldots,\widetilde{u}_{m}(t))$, where $\widetilde{u}_{i}(t)$ is defined as follows:
$$\widetilde{u}_{i}(t)=\begin{cases}
u_{i}(t),{\rm if} \; a_{i}(t)=0; \\
\neg u_{i}(t),{\rm else.}
\end{cases}$$
Furthermore, we denote the action with no attack on any control node as $\widehat{a}=(0,0,\ldots,0)$. The goal of the attacker is to select the attack time point and the control node to be attacked so that the state of the PBCN reaches the target state $X_d$ after time $T$, where $X_{d}$ can be chosen arbitrarily, as long as there is a sequence of controls $U_{T}$ such that $Pr\{X(T;X(0);U_{T})=X_{d}\}=1$. The main problems we are interested in consist of the following:
\begin{enumerate}
    \item  How to choose the attack time points and the control nodes so that the state of the PBCN reach the target state $X_d$ after time $T$ with probability 1?
    \item On the basis of the previous problem, how to minimize the expectation of time points that need to be attacked?
    \item On the basis of the first problem, how to minimize the expectation of the number of attacks on the control node?
\end{enumerate}

In some literature that use RL methods to solve the control problems on the PBCN, the state of the PBCN can be naturally set to be the state of the MDP. However, the objectives of these three problems are related to the time $T$,  so we need the states of the PBCN along with the time $t$ to be the states of MDP. Under this setting, the state set $\mathcal{S}$ of the MDP is $\mathcal{B}^{n} \times \{0,1,\ldots,T\}$, and 
the terminal states are $s = (X(T),T) \in \mathcal{B}^{n} \times \{T\}$. The reward function $\mathcal{R}$ of the MDP for each of the three problems are defined as follows\textcolor{blue}{.} for problem 1), 
\begin{equation}
 R_{t+1}=R(s^{\prime}=(X(t+1),t+1),s=(X(t),t),a)=0
\end{equation}
for all $t \in \{0,1,\ldots,T-2\}$, $a \in \mathcal{B}^{m}$, and:
\begin{equation}
    \begin{cases}
 R_{T}=R(s^{\prime}=(X(T),T),s=(X(T-1),T-1),a)\\
 =1, \; {\rm if}\; X(T)=X_{d}, \\ 
 R_{T}=R(s^{\prime}=(X(T),T),s=(X(T-1),T-1),a)\\
 =0,\; {\rm else}. \\
\end{cases} 
\end{equation}
For problem 2), 
\begin{equation}
    \begin{cases}
    R_{t+1}=R(s^{\prime}=(X(t+1),t+1),s=(X(t),t),a)\\ 
    =0,\; {\rm if} \; a=\widehat{a},\\
 R_{t+1}=R(s^{\prime}=(X(t+1),t+1),s=(X(t),t),a)\\
 =r^{\prime}, \;{\rm else}, \\ 
\end{cases} 
\end{equation}
for all $t \in \{0,1,\ldots,T-2\}$, $a \in \mathcal{B}^{m}$, and:
\begin{equation}
    \begin{cases}
    R_{T}=R(s^{\prime}=(X(T),T),s=(X(T-1),T-1),a)\\ 
    =1,\; {\rm if} \; a=\widehat{a} \; {\rm and} \; X(T)=X_{d},\\
    R_{T}=R(s^{\prime}=(X(T),T),s=(X(T-1),T-1),a)\\ 
    =1+r^{\prime},\; {\rm else \; if \;} a \neq \widehat{a} \; {\rm and} \; X(T)=X_{d},\\
 R_{T}=R(s^{\prime}=(X(T),T),s=(X(T-1),T-1),a)\\ 
    =0,\; {\rm else \; if \;} a =\widehat{a} \; {\rm and} \; X(T) \neq X_{d},\\
 R_{T}=R(s^{\prime}=(X(T),T),s=(X(T-1),T-1),a)\\ 
    =r^{\prime},\; {\rm else}.\\
\end{cases} 
\end{equation}
For problem 3), 
\begin{equation}
\begin{split}
    R_{t+1}&=R(s^{\prime}=(X(t+1),t+1),s=(X(t),t),a)\\
 &=r^{\prime}\cdot HW(a)
\end{split}
\end{equation}
for all $t \in \{0,1,\ldots,T-2\}$, $a \in \mathcal{B}^{m}$, and:
\begin{equation}
    \begin{cases}
    R_{T}=R(s^{\prime}=(X(T),T),s=(X(T-1) 
    ,T-1),a)\\=1+r^{\prime}\cdot HW(a),\; {\rm if} \; X(T)=X_{d},\\
 R_{T}=R(s^{\prime}=(X(T),T),s=(X(T-1)
    ,T-1),a)\\=r^{\prime}\cdot HW(a),\; {\rm else},\\
\end{cases} 
\end{equation}
where $HW(a)=\sum_{i=1}^{m}a_{i}$ is the Hamming weight of the action $a$. In addition, choosing an appropriate value of $r^{\prime}<0$ in equations (12)-(15) ensures if $Pr\{X_{A_T}(T)=X_d\}=1$ and $Pr\{X_{A^{\prime}_T}(T)= X_d\} \neq 1$, then $\mathbb{E}[G(A_T)]> \mathbb{E}[G(A^{\prime}_T)]$, where $A_T$ is the attack policy, $\mathbb{E}[G(A_T)]$ is the expectaction of the total reward under $A_T$, and $X_{A_T}(T)$ is the state of PBCN at time $T$ under $A_T$. Next, we analyze the value range for $r^{\prime}$ as follows. Since $T$ is a fixed value and the primary objective for all three problems is to guide the PBCN to the taget state, it is reasonable to choose $\gamma=1$. Then for problem 2), we have:
\begin{equation}
    \begin{cases}
    \min\limits_{Pr\{X_{A_T}(T)=X_d\}=1}(\mathbb{E}[G(A_T)]) \geq 1+Tr^{\prime} \textcolor{blue}{,}\\
    \max\limits_{Pr\{X_{A_T}(T)=X_d\} \neq 1}(\mathbb{E}[G(A_T)]) \leq 0 \textcolor{blue}{,}\\
\end{cases} 
\end{equation}
and for problem 3):
\begin{equation}
    \begin{cases}
        \min\limits_{Pr\{X_{A_T}(T)=X_d\}=1}(\mathbb{E}[G(A_T)]) \geq 1+Tmr^{\prime} \textcolor{blue}{,}\\
        \max\limits_{Pr\{X_{A_T}(T)=X_d\} \neq 1}(\mathbb{E}[G(A_T)]) \leq 0 \textcolor{blue}{,}\\
\end{cases} 
\end{equation}
So, $r^{\prime}>-\frac{1}{T}$ for problem 2) and $r^{\prime}>-\frac{1}{mT}$ for problem 3).
By representing the attacked PBCN model in the MDP framework described as above, the optimal false data injection attack problems 1), 2) and 3) can be formulated as finding the strategy $A_{T}^{*}$ such that:

\begin{equation}
    A_{T}^{*}=\mathop{{\rm argmax}}\limits_{A_{T}}\mathbb{E}_{A_{T}}\big[\sum_{t=0}^{T-1}\gamma^{t}R_{t+1} |s(0) = (X(0), 0) \big].
\end{equation}

\section{Algorithms for Optimal Attack Strategies of a PBCN}
In this section, we introduce the QL algorithm, and give an improved QL algorithm and a modified QL algorithm to solve the optimal false data injection attack problem for the PBCN (9).
\subsection{QL Algorithm for Optimal Attack on PBCN}
In order to provide the attacker's attack strategy in a model-free framework, we can use the QL algorithm. QL is a model-free, off-policy temporal difference control algorithm. The goal of the QL is to find the optimal action-value function $q_{*}$ and the action-value function is updated by the following formula:
\begin{equation}
    q(S_{t},A_{t}) \! \leftarrow \! q(S_{t},A_{t}) + \alpha [R_{t+1}+\gamma \mathop{{\rm max}}\limits_{a}q(S_{t+1},a)-q(S_{t},A_{t})] \textcolor{blue}{.}
\end{equation}

Under the assumption that all the state-action pairs be visited infinitely often and a variant of the usual stochastic approximation conditions on the sequence of step-size parameters $\alpha$, QL has been shown to converge to the optimal action-value function $q_{*}$ with probability one \cite{watkins1992q}.
The QL algorithm is shown in Algorithm 1.
\begin{remark}
Algorithm 1 can solve the three interested problems. The only difference is that the reward is set in different ways. Similarly, Algorithms 2 and 3 in the sequel also can solve the three interested problems of different reward settings. 
\end{remark}
\begin{algorithm}\small
\renewcommand{\algorithmicrequire}{\textbf{Input:}}
\renewcommand{\algorithmicensure}{\textbf{Output:}}
\caption{The QL algorithm for optimal attack on PBCN}
\begin{algorithmic}[1]
\REQUIRE discount rate $\gamma \in (0,1]$, step size $\alpha \in (0,1]$,  $\epsilon>0$ is sufficiently small.
\ENSURE action-value function $q(s,a)$
\STATE Initialize $q(s,a)$ for all $s \in \mathcal{B}^{n} \times \{0,1,\ldots,T\},\; a \in \mathcal{A}(s)$, arbitrarily except that $q({\rm terminal},\cdot)=0$
\FOR{each episode}
\STATE Initialize $s=(X(0),0)$
\FOR{$t=0$ to $T-1$}
\STATE Choose action $a$ according to $s$ using policy derived from $q$, e.g., by $\epsilon-{\rm greedy}$ policy, that is:
$$a \leftarrow   \begin{cases}
 \mathop{\rm argmax}\limits_{a}q(s,a) & {\rm with \; probability} \; 1-\epsilon \\ 
 {\rm a \; random \; action} & {\rm with \; probability} \; \epsilon 
\end{cases} $$ 
\STATE Take action $a$, observe $R_{t+1}$ and $s^{\prime}=(X(t+1),t+1)$
\STATE $q(s,a) \! \leftarrow \! q(s,a) + \alpha [R_{t+1}+\gamma \mathop{{\rm max}}\limits_{a}q(s^{\prime},a)-q(s,a)]$
\STATE $s \leftarrow s^{\prime} $
\ENDFOR
\ENDFOR
\end{algorithmic}
\end{algorithm}

\subsection{Improved QL Algorithm for Optimal Attack on PBCN}
Note that the QL algorithm creates a look-up table for all state-action pairs, thus requiring a
large amount of memory when the state-action space is very large, as the state-action space
grows exponentially with the number of nodes and inputs. When the state-action space
becomes very large, the basic QL algorithm is often unable to handle this situation due to
the dramatic increase in required memory. In attack problems for a PBCN, when the state-
action space becomes large, there may also be many state-action pairs that are not accessed.
Let $N$ denote the number of state-action pairs that can be accessed. Then, in the attacked
PBCN model, the space requirement for the action-value function $q(s,a)$ is $2^{n+m} \cdot
T$. We propose a small modification to Algorithm 1 by recording and updating the
corresponding value function only when the state-action pair is encountered for the first
time, such that it can be used for attack problems in large-scale PBCNs if $N\ll2^{n+m}
\cdot T$. Under this setting, the space requirement for the action-value function $q(s,a)$ is
reduced from $2^{n+m} \cdot T$ in Algorithm 1 to $N$ in Algorithm 2.
In addition, note that a condition for the convergence of the QL algorithm is that all state-action pairs be visited infinitely often, which cannot be satisfied in actual processes.
Therefore, the policy obtained by QL after a finite number of episodes may not be optimal,
but the learning process may have experienced the optimal state-action sequence. One approach to address this is by recording the total rewards of policies encountered during the learning process for each initial state. After the learning process, we compare the total rewards of policies obtained by QL with the recorded optimal policies and select the policy with the higher expected total reward as our final policy. By combining these two methods, we obtain an improved QL algorithm shown
in Algorithm 2.

\begin{remark}
In Algorithm 2, we still use the same updated function (21) of the QL. Therefore, Algorithm 2 can still converge to the optimal policy with probability 1 when the conditions in the QL algorithm are satisfied. 
\end{remark}
\begin{remark}
In Algorithm 2, $max\_A$ is a list that used to record the best action sequences encountered during the learning process and $max\_r$ is the corresponding return. 
\end{remark}

\begin{remark}
In Algorithm 2, $M<0$ is a constant that serves as the initial value for $max_r$. Since $M<0$ needs to be small enough such that the total reward of any policy is larger than it, $M$ can be any number as long as the condition $M<-\gamma^{T-1}$ is satisfied.
\end{remark}

\begin{remark}
    In Algorithm 2, $Q$ is a dict of action values, where the elements are key-value pairs.
    The key is the state, and the value is the corresponding action value in that state. $Q.keys()$
    represents the set of states that have been recorded in $Q$.
\end{remark}

\begin{remark}
    Since Algorithm 2 is based on QL, the agent must choose the optimal action-value from the
    $2^m$ attack actions at each time step. However, if we use the DQN algorithm, the agent
    not only has to select the optimal action value from the $2^m$ attack actions but also update
    the network parameters, resulting in higher computational complexity compared to Algorithm 2.
    Moreover, Algorithm 2 provides  better convergence guarantees than DQN.
\end{remark}

\begin{algorithm}\small
\renewcommand{\algorithmicrequire}{\textbf{Input:}}
\renewcommand{\algorithmicensure}{\textbf{Output:}}
\caption{The modified QL algorithm for optimal attack on PBCN}
\begin{algorithmic}[1]
\REQUIRE discount rate $\gamma \in (0,1]$, step size $\alpha \in (0,1]$, $M<0$ and $\epsilon>0$ are sufficiently small.
\ENSURE action-value function $Q$, $max\_r$, $max\_A$
\STATE Initialize $Q \leftarrow$ an empty dict, $Visited\_R \leftarrow$ an empty dict, $max\_r=M$, $max\_A \leftarrow$ an empty list.
\FOR{each episode}
\STATE Initialize $s=(X(0),0)$, $temp\_r=0$, $temp\_A \leftarrow$ an empty list
\IF{$s \; not \; in \; Q.keys(\;)$}
\STATE Initialize $q(s,a)$ for all $a \in \mathcal{A}(s)$ arbitrarily.
\STATE $Q(s) \leftarrow q(s,a) $
\ELSE
\STATE $q(s,a) \leftarrow Q(s)$
\ENDIF
\FOR{$t=0$ to $T-1$}
\STATE Choose action $a$ according to $s$ using policy derived from $q$, e.g., by $\epsilon-{\rm greedy}$ policy, that is:
$$a \leftarrow   \begin{cases}
 \mathop{\rm argmax}\limits_{a}q(s,a) & {\rm with \; probability} \; 1-\epsilon \\ 
 {\rm a \; random \; action} & {\rm with \; probability} \; \epsilon 
\end{cases} $$ 
\STATE Take action $a$, observe $R_{t+1}$ and $s^{\prime}=(X(t+1),t+1)$
\IF{$s^{\prime} \; not \; in \; Q.keys(\;)$}
\STATE Initialize $q(s^{\prime},a)$ for all $a \in \mathcal{A}(s^{\prime})$ arbitrarily \STATE $Q(s^{\prime}) \leftarrow q(s^{\prime},a) $
\ELSE
\STATE $q(s^{\prime},a) \leftarrow Q(s^{\prime})$
\ENDIF
\STATE $q(s,a) \! \leftarrow \! q(s,a) + \alpha [R_{t+1}+\gamma \mathop{{\rm max}}\limits_{a}q(s^{\prime},a)-q(s,a)]$
\STATE $Q(s) \leftarrow q(s,a) $
\STATE $s \leftarrow s^{\prime} $
\STATE $temp\_r \leftarrow  temp\_r + \gamma^{t}R_{t+1}$
\STATE put action $a$ into $temp\_A$
\ENDFOR
\IF{$temp\_A \; not \; in \; Visited\_R.keys(\;)$}
\STATE $N=1$, $average=temp\_r$
\STATE $Visited\_R(temp\_A) \leftarrow (N, average)$
\ELSE
\STATE $(N, average)=Q(temp\_A)$
\STATE $Visited\_R(temp\_A) \leftarrow (N+1, average + \frac{temp\_r -average}{N+1})$
\ENDIF
\ENDFOR
\FOR{$A$ in $Visited\_R.keys(\;)$}
\STATE $(N\_A, average\_A)=Visited\_R(A)$
\IF{$average\_A > max\_r$}
\STATE $max\_r \leftarrow average\_A$
\STATE $max\_A \leftarrow A$
\ENDIF
\ENDFOR
\end{algorithmic}
\end{algorithm}

\subsection{Computational Complexity}
We analyze the time and space complexity of our proposed algorithms as follows. In all the proposed
algorithms, the agent must choose the optimal action-value from the $2^m$ attack actions
at each time step, resulting in a time complexity of $O(2^m)$. In each episode out of $M$
episodes, this operation needs to be be performed $T$ times. Thus, the total time complexity of
Algorithm 1 is $O(MT2^m)$, while Algorithm 2 includes an additional operation to find the maximum value, resulting in a total time complexity of $O(MT2^m + M)$. Now, we consider space
complexity. Since it is necessary to store the values of state-action pairs, the space
complexity involved here for Algorithm 1 is $O(T2^{n+m})$. The space complexity of
Algorithm 2 for storing the values of state-action pairs is $O(N2^m)$, where $N$ is the
number of state-action pairs that can be accessed. In addition to storing the values of state-action pairs, it is necessary to record the expected total reward of policies for Algorithm 2,
resulting in a space complexity of $O(N2^m + M)$.

\section{Simulation Results}

In this section, we show how to solve the three problems on attacked PBCN by the results obtained in the previous section. We consider two PBCN models, one with 10 nodes, 3 control inputs and another one with 28 nodes, 3 control inputs to prove the efficiency of the proposed algorithms. Since $T$ is not a very critical parameter in the simulation, we can actually choose $T$ as any number. However, taking the value of $T$ too large requires a lot of computation, so we choose $T$ as a small number. For the following two examples, we choose $T=5$.

\subsection{Example 5.1}

Consider the following PBCN model with 10 nodes and 3 control inputs, which is is similar to the model of the lactose operon in the Escherichia Coli derived from \cite{veliz2011boolean}:
\begin{align}
   x_{1}(t+1)&=x_{4}(t) \land \neg x_{5}(t) \land \neg x_{6}(t) \notag\\
   x_{2}(t+1)&=x_{3}(t+1)=x_{1}(t) \notag\\
   x_{4}(t+1)&=\neg u_{1}(t), x_{5}(t+1)=\neg x_{7}(t) \land \neg x_{8}(t) \notag\\
   x_{6}(t+1)&=\neg x_{7}(t) \land \neg x_{8}(t) \vee x_{5}(t) \notag\\
   x_{7}(t+1)&=x_{3}(t) \land x_{9}(t),  x_{8}(t+1)=x_{9}(t) \vee x_{10}(t) \notag\\
   x_{9}(t+1)&=\begin{cases}
    x_{2}(t) \vee (u_{3}(t) \land \neg u_{1}(t)), \; p=0.6\notag\\
    x_{9}(t), \; p=0.4
   \end{cases}\notag\\
   x_{10}(t+1)&=((u_{2}(t) \land x_{2}(t)) \vee u_{3}(t)) \land \neg u_{1}(t)
\end{align}
and we choose the state (0, 0, 0, 0, 1, 1, 0, 0, 0, 0) of the PBCN to be the initial state of the PBCN. The attacker aims to change the state of PBCN to the desired state (0, 0, 0, 1, 1, 1, 0, 0, 0, 0) at $T=5$ by tampering with the control inputs. Moreover, we choose $r^{\prime}=-0.1$ and $r^{\prime}=-0.05$ for problems 2), 3), respectively. As for the step size $\alpha$ and the discount rate $\gamma$, we choose $\gamma=1$ for all three problems, and $\alpha=0.01$ for problem 1 and problem 2 while $\alpha=0.09$ for problem 3. By following Algorithms 1 and 2, we obtain optimal attacks on the PBCN (20). We run 1000
independent experiments to obtain the average reward for each episode for these three
problems, and the results are shown in Figure 2. As can be seen in Figure 2, the average
reward increases with the number of episodes and eventually converges, implying that an
optimal attack policy has been obtained. By running 100 independent experiments for
different numbers of episodes, we obtain the relationship between the probability of
learning the optimal attack policy and the number of episodes for both algorithms. As
shown in Figure 3, the probability of learning the optimal attack policy increases with the number of episodes and eventually converges to 1 for both algorithms. However, under the same number of episodes, the probability of learning the optimal policy using Algorithm 2 is higher than that of Algorithm 1, implying that our proposed improved QL algorithm is more effective than traditional QL in solving the optimal attack problem for a PBCN.

\begin{figure}[H]
\centerline{\includegraphics[scale=0.5]{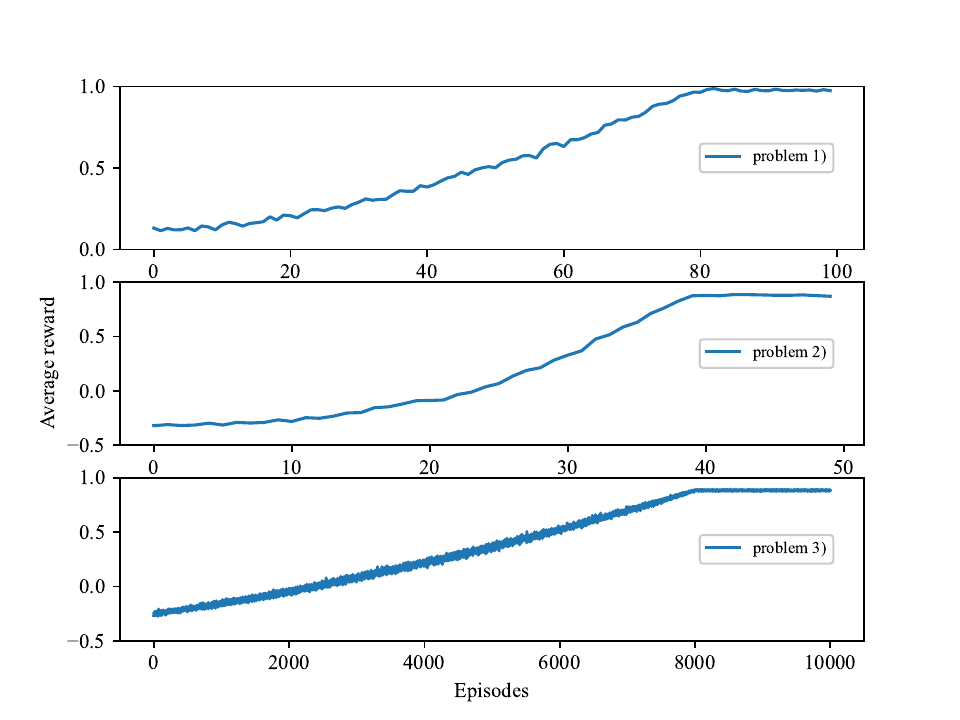}}
\caption{For system (22), the average total rewards of 1000 independent experiments in problem 1), 2) and 3). }
\label{fig2}
\end{figure}

\begin{figure}[H]
\centerline{\includegraphics[scale=0.5]{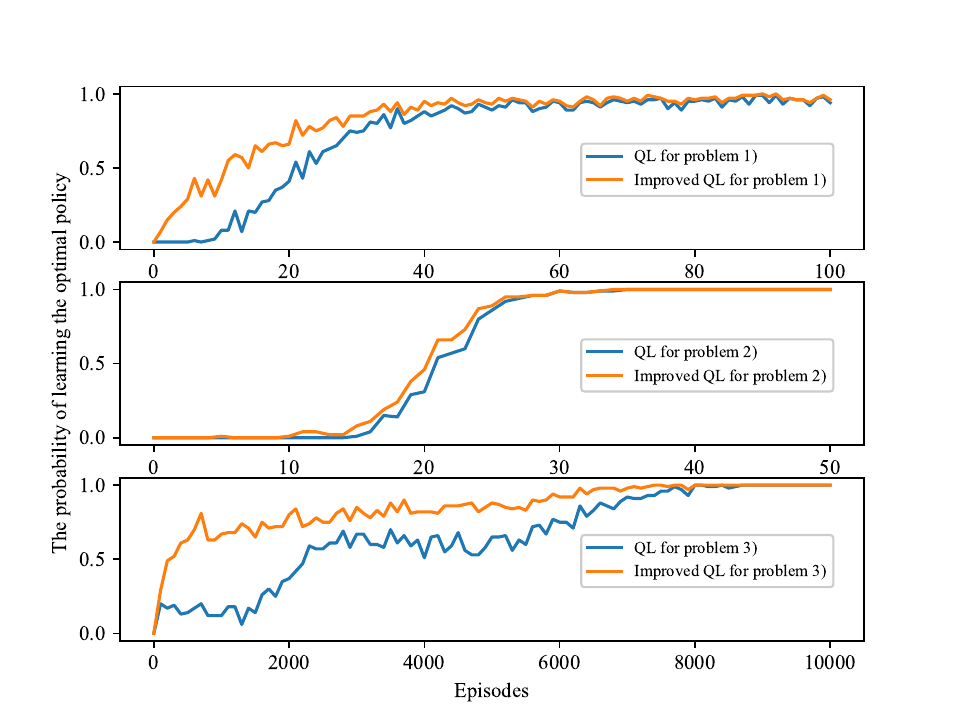}}
\caption{For system (22), the comparison of the probabilities of learning the optimal policy between Algorithm 1 (QL) and Algorithm 2 (improved QL) in problem 1), 2) and 3).}
\label{fig3}
\end{figure}

\subsection{Example 5.2}

Consider the following PBCN model with 28 nodes and 3 control inputs \cite{acernese2020double}, which is a reduced-order model of 32-gene T-cell receptor kinetics model given in \cite{zhang2020efficient}:
\begin{align}
    x_{1}(t+1)&=x_{6}(t) \land x_{13}(t), x_{2}(t+1)=x_{25}(t) \notag\\
    x_{3}(t+1)&=x_{2}(t), x_{4}(t+1)=x_{28}(t) \notag\\
    x_{5}(t+1)&=x_{21}(t), x_{6}(t+1)=x_{5}(t) \notag\\
    x_{7}(t+1)&=(x_{15}(t) \land u_{2}(t)) \vee (x_{26}(t) \land u_{2}(t)) \notag\\
    x_{8}(t+1)&=x_{14}(t), x_{9}(t+1)=x_{18}(t) \notag\\
    x_{10}(t+1)&=x_{25}(t) \land x_{28}(t), x_{11}(t+1)=\neg x_{9}(t) \notag\\
    x_{12}(t+1)&=x_{24}(t), x_{13}(t+1)=x_{12}(t) \notag\\
     x_{14}(t+1)&=x_{28}(t) \notag\\
    x_{15}(t+1)&=(\neg x_{20}(t)) \land u_{1}(t) \land u_{2}(t) \notag\\
    x_{16}(t+1)&=x_{3}(t), x_{17}(t+1)=\neg x_{11}(t) \notag\\
    x_{18}(t+1)&=x_{2}(t) \notag\\
    x_{19}(t+1)&=(x_{10}(t) \land x_{11}(t) \land x_{25}(t) \land x_{28}(t)) \vee (x_{11}(t) \land \notag\\
    & \quad \; x_{23}(t) \land x_{25}(t) \land x_{28}(t)) \notag\\
    x_{20}(t+1)&=x_{7}(t) \vee \neg x_{26}(t) \notag\\
    x_{21}(t+1)&=x_{11}(t) \vee x_{22}(t) \notag\\
    x_{22}(t+1)&=x_{2}(t) \land x_{18}(t), x_{23}(t+1)=x_{15}(t) \notag\\
    x_{24}(t+1)&=x_{18}(t), x_{25}(t+1)=x_{8}(t) \notag\\
    x_{26}(t+1)&=\begin{cases}
    \neg x_{4}(t) \land u_{3}(t), \; p=0.5; \notag\\
     x_{26}(t), \; p=0.5; \notag\\
    \end{cases} \notag\\
    x_{27}(t+1)&=x_{7}(t) \vee (x_{15}(t) \land x_{26}(t)) \notag\\
    x_{28}(t+1)&=\neg x_{4}(t) \land x_{15}(t) \land x_{24}(t)
\end{align}

and we choose the state (0, 0, 0, 0, 1, 1, 1, 0, 0, 0, 1, 0, 0, 0, 0, 0, 0, 0, 0, 1, 1, 0, 0, 0, 0, 1, 1, 0) of the PBCN to be the initial state of the PBCN, and the desired state is (0, 0, 0, 1, 1, 1, 1, 0, 0, 0, 1, 0, 0, 1, 0, 0, 0, 0, 0, 1, 1, 0, 0, 0, 0, 1, 1, 0).
Moreover, we set $T=5$ for all the three problems, and we choose $r^{\prime}=-0.1$ and  $r^{\prime}=-0.05$ for problems 2) and 3), respectively. As for the step size $\alpha$ and the discount rate $\gamma$, we choose $\gamma=1$ and $\alpha=0.05$ for all three problems. By following the improved QL algorithm (Algorithm 2), we can also obtain the optimal attack policy for large scale PBCN, which cannot be achieved by directly applying the traditional QL algorithm (Algorithm 1). Similar to example 5.1, the average reward on each episode for these three problems are shown in Figure 4, where one can see that the average reward increases with the number of episodes and eventually converges. Figure 5 also shows the property that the probability of learning the optimal attack policy increases with the number of episodes and eventually converges to 1.

\begin{figure}[!htbp]
\centerline{\includegraphics[scale=0.5]{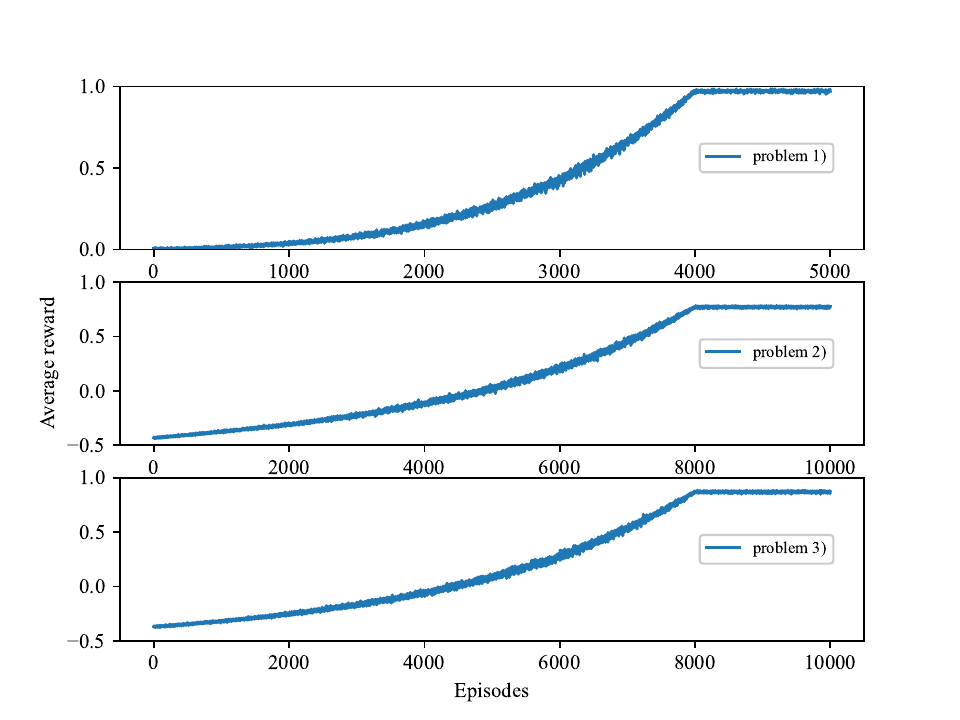}}
\caption{For system (23), the average total rewards of 1000 independent experiments in problem 1), 2) and 3).}
\label{fig4}
\end{figure}

\begin{figure}[!htbp]
\centerline{\includegraphics[scale=0.5]{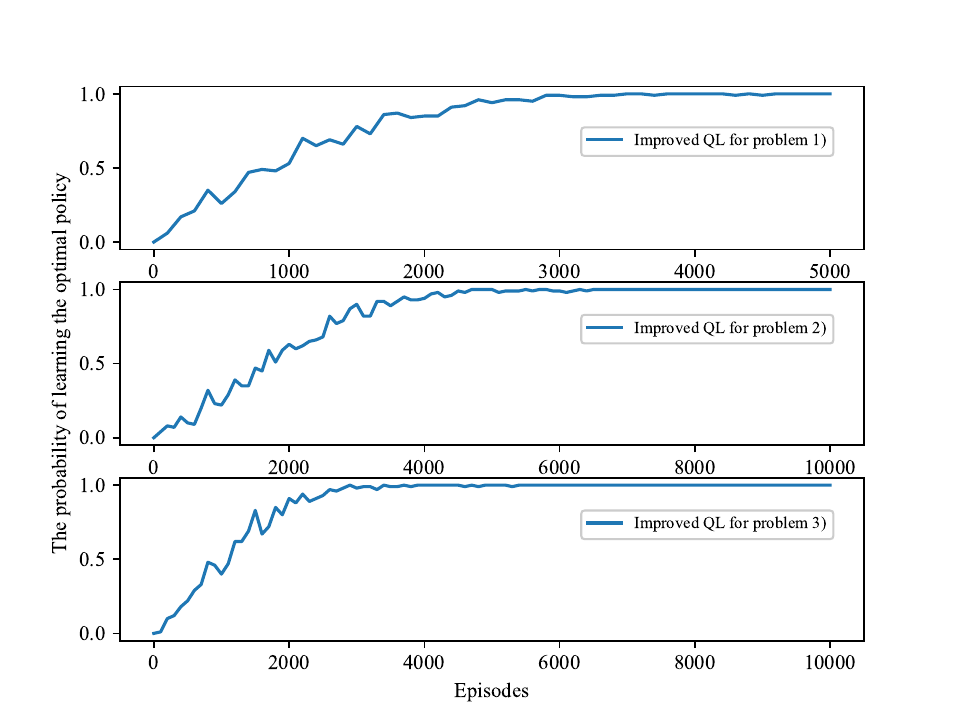}}
\caption{For system (23), the probabilities of learning the optimal policy using Algorithm 2(improved QL) in problem 1), 2) and 3).}
\label{fig5}
\end{figure}

\section{Conclusion}
In this paper, we have investigated optimal false data injection attack problems on PBCNs
using model-free RL methods. We have proposed an improved QL algorithm to solve these
problems more effectively. Additionally, our proposed algorithm can solve the optimal attack
problem for some large-scale PBCNs that the standard QL algorithm cannot handle. We have
provided two examples to illustrate the validity of our proposed algorithms. However, we
have only discussed false data injection attack problems from the attacker’s perspective. In
the future, we may study resilient control problems against the false data injection attacks
proposed in this paper, such that the system can remain stable through controller design,
making it more robust. 

\bibliographystyle{ieeetr}

\bibliography{root}

\end{document}